# Record Surface State Mobility and Quantum Hall Effect in Topological Insulator Thin Films via Interface Engineering

*Nikesh Koirala*[‡§], *Matthew Brahlek*[‡§†], *Maryam Salehi*[#§], *Liang Wu*[‖], *Jixia Dai*[‡], *Justin Waugh*[⊥], *Thomas Nummy*[⊥], *Myung-Geun Han*[¶], *Jisoo Moon*[‡], *Yimei Zhu*[¶], *Daniel Dessau*[⊥], *Weida Wu*[‡], *N. Peter Armitage*[‖], *Seongshik Oh*[‡*]

§ *These authors contributed equally to this work*

[‡]Department of Physics & Astronomy, Rutgers, The State University of New Jersey, Piscataway, New Jersey 08854, U.S.A.

[#]Department of Materials Science and Engineering, Rutgers, The State University of New Jersey, Piscataway, New Jersey 08854, U.S.A.

[‖]Department of Physics & Astronomy, The Johns Hopkins University, Baltimore, Maryland 21218, U.S.A.

[⊥]Department of Physics, University of Colorado, Boulder, Colorado 80309, U.S.A

[¶]Condensed Matter Physics & Materials Science, Brookhaven National Lab, Upton, NY 11973, U.S.A.

ABSTRACT: Material defects remain as the main bottleneck to the progress of topological insulators (TIs). In particular, efforts to achieve thin TI samples with dominant surface transport have always led to increased defects and degraded mobilities, thus making it difficult to probe the quantum regime of the topological surface states. Here, by utilizing a novel buffer layer scheme composed of an $In_2Se_3/(Bi_{0.5}In_{0.5})_2Se_3$ heterostructure, we introduce a quantum generation of $Bi_2Se_3$ films with an order of magnitude enhanced mobilities than before. This scheme has led to the first observation of the quantum Hall effect in $Bi_2Se_3$.

Although numerous novel electronic phenomena have been observed in topological insulators (TIs) over the past years,[1,2,3,4,5] material defects have been a major obstacle hindering further progress toward many applications of TIs. In the highest quality, undoped TI crystals, bulk defects dominate the transport,[6,7] and while efforts to compensate-dope and combinations of anion or cation mixing have helped suppress bulk conduction, such efforts simultaneously degrade carrier mobility as well.[8,9] Further, when the samples are made thin, either by mechanical exfoliation or by thin film growth, surface defects become an additional source of carriers, which lead to parasitic accumulation layers.[10,11,12,13,14,15,16,17] Similar to bulk samples, compensation doping in thin films can decrease the bulk contribution, but only at the cost of degraded topological surface state (TSS) mobility.[2] Considering that both high mobilities and low carrier densities are essential to revealing quantum nature of any materials, a solution to this problem will pave a way to the age of quantum TIs.

It is natural to postulate that a chemically and structurally matched substrate should simultaneously suppress both interfacial and bulk defects, which are responsible for low mobilities and high carrier densities in TI films. At the moment, such a single-crystal substrate does not exist, with the current selections either increasing bulk defects or creating interfacial defects. [12,13,14,15,16,17] However, by starting with a standard substrate, molecular beam epitaxy (MBE) does allow a virtual-substrate composed of an electrically insulating buffer layer to be engineered on the atomic-scale.[18,19] Here we show that an isostructural buffer layer consisting of 20 quintuple layers (QL, 1 QL ≈ 1 nm) $In_2Se_3$ followed by 20 QL $(Bi_{0.5}In_{0.5})_2Se_3$ (referred to as BIS-BL herein) grown on an $Al_2O_3$ substrate indeed serves as an ideal template for high quality $Bi_2Se_3$ growth with minimal interfacial and bulk defects.[20] This suppression of material defects has led to low carrier density $Bi_2Se_3$ films with the highest reported TSS mobility, which culminated in our observation of the quantum Hall effect (QHE).

$Bi_2Se_3$, $In_2Se_3$ and the solid-solution $(Bi_{0.5}In_{0.5})_2Se_3$ all share the same layered structure with covalently bonded QLs, which are held together by the weak van der Waal's (vdW) force.[19,20] Due to vdW bonding, $Bi_2Se_3$ can be grown on substrates with large lattice mismatch since the film relaxes to its bulk lattice constant within the first QL. This allows the defects formed at the interface to be isolated within the first layer and subsequent layers have substantially suppressed defects even when grown on conventional substrates with large lattice mismatch such as $Al_2O_3(0001)$ (~14% lattice mismatch).[12,15] Additionally, when $Bi_2Se_3$ is grown on top of a substrate with a better lattice match, such as Si(111) and InP(111) (~8% and ~0.2%, respectively), interfacial defects are suppressed, but transport measurements show that additional bulk defects are created, which are likely due to the more reactive substrate surface. [14,16,21] $In_2Se_3$ and $(Bi_{0.5}In_{0.5})_2Se_3$ with their layered structure minimize strong chemical bonding at the interface and with the small lattice mismatch (3.3% and 1.6% respectively) to $Bi_2Se_3$, serve as ideal candidates to create a buffer layer for $Bi_2Se_3$.[19,20] As detailed next, we make use of these properties of $In_2Se_3$ and $(Bi_{0.5}In_{0.5})_2Se_3$ to create a high quality virtual-substrate for $Bi_2Se_3$.

Figure 1a shows a schematic representation of the growth procedure of $Bi_2Se_3$ grown on 20 QL $In_2Se_3$ – 20 QL $(Bi_{0.5}In_{0.5})_2Se_3$ buffer layer (See Supporting Information Part I). Unlike $Bi_2Se_3$, $In_2Se_3$ has at least three phases,[22] and when deposited directly onto the poorly lattice matched $Al_2O_3$ substrate it grows in a disordered form. Therefore, to grow a high-quality, single-phase $In_2Se_3$ layer requires an initial seed layer of 3 QL $Bi_2Se_3$ that is deposited at 135 °C; this serves as a template for the 20-QL-thick $In_2Se_3$ layer to be deposited at 300 °C. At this stage the underlying 3 QL of $Bi_2Se_3$ remains conducting, which is undesirable for transport studies of the main $Bi_2Se_3$ layer to be grown later. In order to make it electrically insulating we heat this entire layer up to 600 °C where the $Bi_2Se_3$ seed layer diffuses through the $In_2Se_3$ and evaporates away, which leaves behind the high quality, insulating $In_2Se_3$ layer directly on the $Al_2O_3$ substrate (See Supporting Information Part II and Figure S1). On top of this a 20-QL-thick insulating $(Bi_{0.5}In_{0.5})_2Se_3$ is then deposited at 275 °C, which acts to suppress In diffusion into the $Bi_2Se_3$ layer (see Supporting Information Part IV).[20,23] This entire structure forms the BIS-BL, and, as we show next, structural probes show that this forms a high quality virtual-substrate for $Bi_2Se_3$ to be deposited at 275 °C.

The reflection high-energy electron diffraction (RHEED) in Figure 1a indicates that the film growth remains two dimensional throughout for the BIS-BL, and results in a flat highly crystalline $Bi_2Se_3$ layer. Further, Figure 1b and 1c show high-angle annular dark-field scanning transmission electron microscopy (HAADF-STEM) images of a 50 QL $Bi_2Se_3$ film indicating highly ordered growth. The most important feature of the films grown on BIS-BL that can be seen in Figure 1c, is the sharp, defect-free interface between $Bi_2Se_3$ and $(Bi_{0.5}In_{0.5})_2Se_3$. This is in contrast to $Bi_2Se_3$ films grown on commonly used substrates such as $Al_2O_3(0001)$ and $Si(111)$, where the interface is more disordered as indicated by STEM/TEM images shown in Figure 1d and 1e, respectively.

Transport measurements are the most sensitive probe to study the presence of defects that supply carriers and cause defect induced scattering. Therefore, to compare the defect density in $Bi_2Se_3$ grown on the BIS-BL to films grown on $Al_2O_3(0001)$ and $Si(111)$ substrates, [12,14] we show in Figure 2a and 2b the sheet carrier density ($n_{sheet}$) and mobility ($\mu$), respectively, extracted from the low magnetic field Hall measurement ($|B| \lesssim 0.5$ T) at $T = 1.5$ K as a function of thickness ($t$) (See Supporting Information Part I). As shown in Figure 2b, the highest mobility for $Bi_2Se_3$ grown on BIS-BL exceeds 16,000 $cm^2V^{-1}s^{-1}$, which is about an order of magnitude larger than the mobility of films grown on $Al_2O_3(0001)$ and $Si(111)$, and this directly shows that BIS-BL significantly suppresses the net defect density. For the entire thickness range of 5 to 60 QL $Bi_2Se_3$ grown on the BIS-BL $n_{sheet} \approx 1 - 3 \times 10^{12}$ $cm^{-2}$, whereas films grown on $Al_2O_3(0001)$ and $Si(111)$ exhibit an order of magnitude larger values of $n_{sheet}$, similarly implying significantly larger defect densities. Further, the thickness independence for films grown on BIS-BL and $Al_2O_3(0001)$ show that the dominant defects come from the interface, whereas films grown on $Si(111)$ have significant thickness dependence ($n_{sheet} \sim t^{1/2}$) implying a combination of both interfacial and bulk defects.[14] Lastly, like films grown on $Al_2O_3(0001)$, some non-linearity in the Hall effect was observed at higher field in films on BIS-BL, indicating multiple conduction channels. However, non-linear Hall fitting can be used to estimate the total carrier density, which is found to be $\lesssim 5 \times 10^{12}$ $cm^{-2}$ over the entire thickness range for BIS-BL, and $\sim 40 \times 10^{12}$ $cm^{-2}$ for films grown on $Al_2O_3$ (see Supporting Information Part III).[12] By simultaneously suppressing both interfacial and bulk defects with this BIS-BL, this is the first time such a large improvement in both $n_{sheet}$ and $\mu$ has been observed in any TI films.

Figure 2 also shows the angle-resolved photoemission spectroscopy (ARPES) spectra of a 30 QL thick $Bi_2Se_3$ films grown on BIS-BL (Figure 2c) and a 50 QL thick $Bi_2Se_3$ grown on $Al_2O_3$ (Figure 2d), both clearly showing the TSS bands, which confirms their non-trivial topology (See Supporting Information Part I and IV and Figure S3). Further, we can compare the position of the surface Fermi levels ($E_F$) of $Bi_2Se_3$ grown on both substrates: For the film grown on $Al_2O_3$, $E_F \approx 0.33$ eV above the Dirac point. At this level, the conduction band is clearly occupied as 2DEG (two-dimensional electron gas) states that form due to downward band bending of bulk conduction band near the surface. In contrast, for the film grown on BIS-BL, $E_F$ is only ~0.17 eV above the Dirac point, which implies that on the surface only the TSS bands are occupied. The lower $E_F$ observed in films grown on BIS-BL compared to films grown directly on $Al_2O_3$ is consistent with the lower carrier density observed in transport measurements.

Having established the TI nature of the films, we now discuss whether the carriers populate the trivial states (bulk or conventional 2DEG) or the TSS. The observation of the surface $E_F$ in the bulk gap from ARPES measurement and low $n_{sheet}$ observed from transport measurement both support that the transport originates from TSS (See Supporting Information Part V and Figure S4).[2,7] This is further supported by the cyclotron resonance (CR) measurement taken from time-domain magneto-terahertz spectroscopy (TDMTS) of two 16 QL films:[24] this measurement provides cyclotron mass ($m^*$), total sheet carrier density ($n_{tot}$) and average mobility ($\mu_{optical}$) (See Supporting Information Part I and VI). To prevent ambient contamination during ~1 day delay between growth (at Rutgers University) and TDMTS measurement (at Johns Hopkins University),[25] one film was capped by ~20 nm Se and another was capped by ~50 nm $MoO_3$.

Sharp CR features were observed in measurements of the complex Faraday rotation (FR) angle. Figure 3a (Figure 3b) shows the real part of Faraday rotation (FR) for a Se- ($MoO_3$-) capped film for different magnetic fields. The inflection point of real part of the FR represents the CR, which shifts to higher frequency with increasing $B$. As shown in Figure 3c for both samples, fitting the FR data provides the CR frequency ($\omega_c$) for each $B$ from which $m^*$ is obtained by using a linear fit $\omega_c = eB/(2\pi m^*)$, where $m_e$ is the bare electron mass (See Supporting Information VII and Figure S5 for imaginary FR and fit). In order to obtain $n_{tot}$ and $\mu_{optical}$, we extracted spectral weight ($\omega_{pD}^2 d$) and scattering rate ($\Gamma_D$) from the Drude–Lorentz fit of the zero field real conductance shown in Figure 3d (See Supporting Information Part VII). We can then get $\mu_{optical} = e/(2\pi \Gamma_D m^*)$ and $n_{tot}$ from:[24]

$$\omega_{pD}^2 d = \frac{n_{tot}\, e^2}{m^* \epsilon_o}. \qquad (1)$$

Additionally, if the carriers populate the TSS, then the Fermi wave-vector ($k_F$) can be obtained solely from spectral weight using:

$$\omega_{pD}^2 d = \frac{k_F(A + 2Bk_F)\, e^2}{2\pi \hbar^2 \epsilon_o}, \qquad (2)$$

where $e$ is the electronic charge, $\epsilon_o$ is the free-space permittivity, and $A = 2.02$ eVÅ and $B = 10.44$ eVÅ$^2$ are the TSS band parameters up to quadratic term (i.e. $E_{TSS} = Ak + Bk^2$) obtained from the fitting.[26] From $k_F$, the sheet carrier density ($n_{TSS}$) and effective mass ($m_{TSS}$) of Dirac-like TSS carriers can be calculated using $n_{TSS} = k_F^2/(2\pi)$ (assuming similar carrier density for two TSSs, which is consistent with Hall measurement; See Supporting Information Part II) and $m_{TSS} = \hbar k_F/v_F$, where $v_F = \frac{dE_F}{\hbar dk}$ is the Fermi velocity of TSS. In Table 1, we list $\omega_{pD}^2 d$, $m^*$, $m_{TSS}$, $n_{tot}$, $n_{TSS}$, $\mu_{optical}$, Hall carrier density ($n_{sheet}$) and Hall mobility ($\mu_{Hall}$) for the two samples (Hall data were obtained from different but nominally identically prepared samples as those used in TDMTS measurement). In addition, Hall data for a fresh uncapped 15 QL thick film is also shown for comparison. For calculation of $\mu_{optical}$, $\Gamma_D$ of ~0.5 THz and ~1 THz were used for Se-capped and MoO$_3$-capped films, respectively. The smaller mobilities obtained for capped samples as compared to uncapped one implies that the capping layers introduce additional surface-scattering sites: in the case of the Se capping, additional carriers as well.

From Table 1 we can see a reduced carrier density for MoO$_3$-capped film as compared to the Se-capped one because MoO$_3$ depletes (n-type) carriers from Bi$_2$Se$_3$ due to its higher electron affinity.[27] The correspondingly smaller $m^* \approx 0.073 m_e$ in MoO$_3$-capped sample compared to $m^* \approx 0.12 m_e$ for Se-capped sample is strong evidence that CR comes from TSS carriers because the effective masses of bulk or 2DEG carriers are carrier density independent (0.11 ~ 0.13$m_e$),[28,29] while the effective mass of Dirac-like carriers in TSS scales with carrier density ($m_{TSS} \propto k_F \propto \sqrt{n_{TSS}}$). Agreement between measured $m^*$ and calculated $m_{TSS}$ gives quantitative evidence of CR originating from TSS carriers. Additionally, agreement between $n_{tot}$, $n_{TSS}$ and $n_{sheet}$ indicate that the TSS conduction accounts for total observed carriers in these films.

Finally, the high mobility of the TSS combined with the extremely low carrier density has allowed us to observe the dissipationless QHE for the first time in Bi$_2$Se$_3$.[5,30] As shown in Figure 4, an 8 QL thick film capped by both MoO$_3$ and Se was hand-patterned into a millimeter-scale Hall-bar and measured in magnetic field up to 34.5 T (inset of Figure 4a). Figure 4a and 4b show the Hall ($R_{Hall}$) and the longitudinal sheet resistance ($R_{sheet}$), respectively, as a function of magnetic field at various temperatures for this film. The sheet carrier density of the film was ~7 $\times$ 10$^{11}$ cm$^{-2}$ as measured from low field Hall slope (< 9 T). The data for 0.3 K shows that $R_{sheet}$ vanishes (0.0 ± 0.5 Ω) above 31 T indicating dissipationless transport, with simultaneous perfect quantization of $R_{Hall}$ = (1.00000 ± 0.00004)$h/e^2$ (25813 ± 1 Ω) above 29 T. Together these indicate that $E_F$ of both the top and bottom surface states have fallen below the first Landau level.[5,30] As shown in Figure 4a and 4b, at the maximum field the quantum-Hall-plateau vanishes between 20 – 50 K, but hints of the QHE persists even up to 70 K.

In conclusion, commonly available substrates are not an ideal growth template for the TI $Bi_2Se_3$, which inevitably suffers from both interfacial and bulk defects. However, the flexibility of MBE has allowed us to engineer an atomic-scale virtual substrate that is tailored to produce high-quality $Bi_2Se_3$ thin films. We have shown that this scheme significantly lowers interfacial and bulk defects, resulting in the highest reported mobility of TSS channels and the first observation of the quantum Hall effect in $Bi_2Se_3$. The development of such high mobility TI films is a major step toward accessing new quantum phenomena, which have been inaccessible to most experimental probes due to parallel bulk conduction and low mobilities.

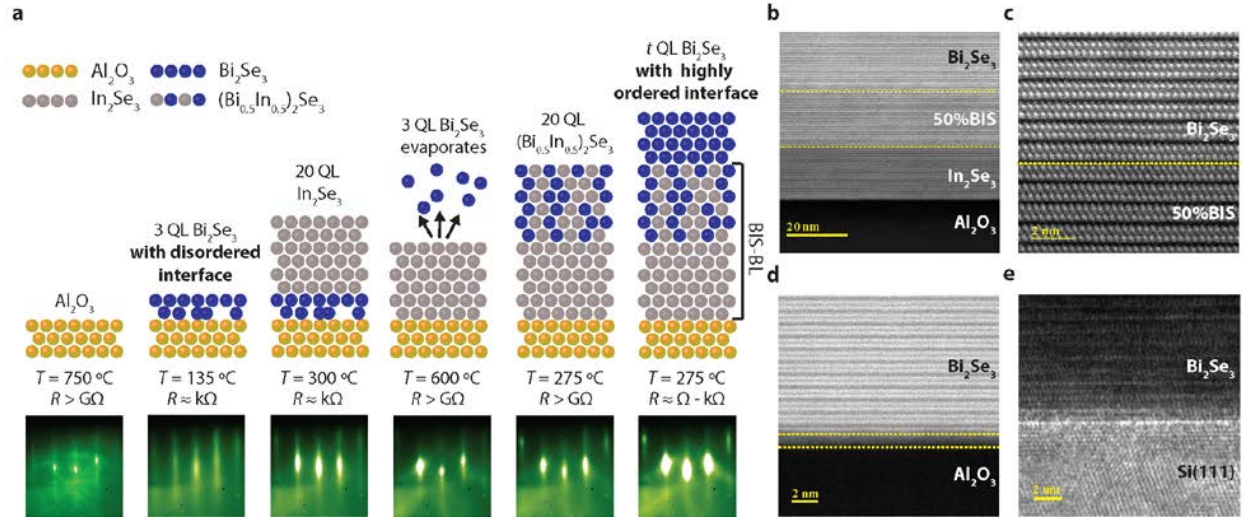

**Figure 1.** Growth process of $Bi_2Se_3$ films on the 20 QL $In_2Se_3$ – 20 QL $(Bi_{0.5}In_{0.5})_2Se_3$ buffer layer (BIS-BL) and comparison with films grown on $Al_2O_3(0001)$ and Si(111). (a) Cartoon showing each stage of film growth along with the corresponding growth temperature ($T$), sheet resistance ($R$) and RHEED images. HAADF-STEM image of $Bi_2Se_3$ grown on (b) BIS-BL, which (c) shows an atomically-sharp interface between $Bi_2Se_3$ and BIS-BL, while (d) $Bi_2Se_3$ grown directly on $Al_2O_3(0001)$ has clearly disordered interface. (e) TEM image of $Bi_2Se_3$ grown on Si(111) (from reference 13). In (b) and (c), $(Bi_{0.5}In_{0.5})_2Se_3$ is written as 50%BIS.

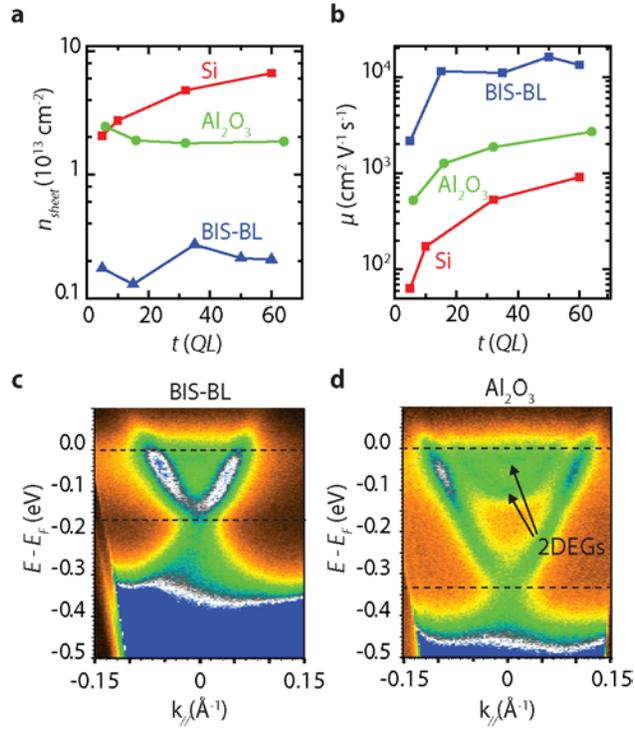

**Figure 2.** Comparison of (a) sheet carrier densities and (b) Hall mobilities of $Bi_2Se_3$ films grown on BIS-BL, $Al_2O_3$(0001) and Si(111) for various film thicknesses. ARPES of $Bi_2Se_3$ grown on (c) BIS-BL and (d) $Al_2O_3$(0001).

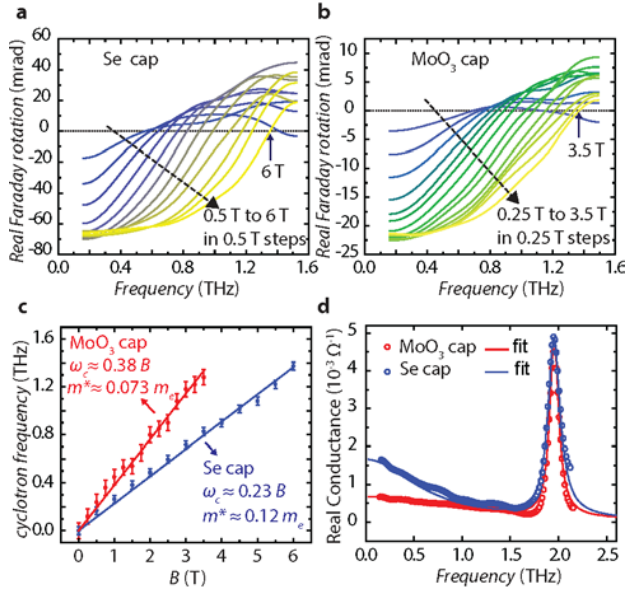

**Figure 3.** Cyclotron resonance and zero field real conductance of two 16 QL thick $Bi_2Se_3$ films grown on BIS-BL and capped by 20 nm Se and 50 nm $MoO_3$ respectively. Real part of complex Faraday rotation at different magnetic fields for (a) Se-capped and (b) $MoO_3$-capped film. For both (a) and (b), dashed arrows indicate the direction of increasing magnetic field. (c) CR frequencies at different magnetic fields for both films. Solid lines are linear fit of $\omega_c = eB/(2\pi m^*)$. (d) Zero field real conductance as a function of frequency along with Drude-Lorentz fit for both films. The peaks near 1.9 THz correspond to bulk phonon mode of $Bi_2Se_3$.

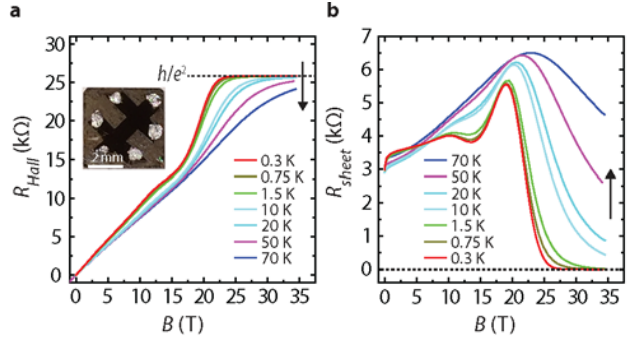

**Figure 4.** Quantum Hall Effect in an 8 QL thick $Bi_2Se_3$ film grown on BIS-BL and capped by both $MoO_3$ and Se. (a) Hall resistance at different temperatures in magnetic field up to 34.5 T, which quantizes to $(1.00000 \pm 0.00004)h/e^2$ $(25813 \pm 1\ \Omega)$ at low temperatures. The inset shows the Hall-bar pattern of the measured film. (b) Corresponding longitudinal sheet resistance, which drops to zero $(0.0 \pm 0.5\ \Omega)$ when Hall resistance quantizes to $h/e^2$. The vertical arrows indicate the direction of increasing temperature.

**Table 1.** Optical and DC transport parameters of two 16 QL thick $Bi_2Se_3$ films grown on BIS-BL with Se and $MoO_3$ capping overlayer along with a 15 QL uncapped film.[a]

| film | $\omega_{pD}^2 d$ (THz$^2$nm) | $m^*/m_e$ | $m_{TSS}/m_e$ | $n_{tot}$ ($10^{12}$ cm$^{-2}$) | $n_{TSS}$ ($10^{12}$ cm$^{-2}$) | $\mu_{optical}$ (cm$^2$V$^{-1}$s$^{-1}$) | $n_{sheet}$ ($10^{12}$ cm$^{-2}$) | $\mu_{Hall}$ (cm$^2$V$^{-1}$s$^{-1}$) |
|---|---|---|---|---|---|---|---|---|
| Se-capped | $2.3 \pm 0.1 \times 10^4$ | $0.12 \pm 0.01$ | $0.12 \pm 0.01$ | $3.5 \pm 0.2$ | $3.4 \pm 0.2$ | 4,700 | 3.1 | 3,600 |
| $MoO_3$-capped | $1.2 \pm 0.1 \times 10^4$ | $0.073 \pm 0.007$ | $0.080 \pm 0.005$ | $1.1 \pm 0.2$ | $1.2 \pm 0.2$ | 3,800 | 1.1 | 3,300 |
| Uncapped | - | - | - | - | - | - | 1.3 | 11,400 |

[a] Spectral density ($\omega_{pD}^2 d$) is obtained from fit of zero field real conductance in TDMTS measurement. Calculated effective mass of TSS ($m_{TSS}$) agrees with cyclotron mass ($m^*$) as do optical sheet carrier densities, $n_{tot}$ and $n_{TSS}$, obtained from equations 1 and 2, respectively, indicating TSS as the origin of cyclotron resonance. A comparison of optical sheet carrier densities and mobility ($\mu_{optical}$) with Hall sheet carrier density ($n_{sheet}$) and Hall mobility ($\mu_{Hall}$), respectively, is also presented. For the uncapped film no optical measurements were performed due to surface aging effect.

AUTHOR INFORMATION

**Corresponding Author**

*Correspondence should be addressed to **ohsean@physics.rutgers.edu**

**Present Addresses**

†Department of Materials Science and Engineering, The Pennsylvania State University, University Park, Pennsylvania 16802, USA

**Author Contributions**

N.K., M.B. and S.O conceived the experiment. N.K., M.B. and M.S. synthesized the thin film samples and performed the transport measurements. N.K. and M.S. carried out the high-field QHE measurement. J.M. assisted the film growth. L.W. and N.P.A performed TDMTS measurements. J.D. and W.W. performed STM measurements. J.W., T.N. and D.D. performed ARPES measurements. M.-Y.G. and Y.Z. performed the TEM measurements. N.K., M.B. and S.O. wrote the manuscript with inputs from other authors. All authors contributed to the scientific analysis and manuscript revisions. All authors have given approval to the final version of the manuscript

**Funding Sources**

Office of Naval Research, National Science Foundation, Gordon and Betty Moore Foundation, U.S. Department of Energy, State of Florida.

ACKNOWLEDGMENT

Thin film growth and transport measurement work (Rutgers University) was supported by ONR (N000141210456), NSF (DMR-1308142) and Gordon and Betty Moore Foundation's EPiQS Initiative (GBMF4418). TDMTS work (JHU) is further supported by Gordon and Betty Moore Foundation (GBMF2628). STM work (Rutgers University) is supported by NSF (DMR-0844807 and DMR-1506618). ARPES work (University of Colorado) is supported by DOE (DE-FG0203ER46066). HAADF-STEM work (BNL) is supported by DOE (DE-AC02-98CH10886). QHE measurement (National High Magnetic Field Laboratory) is supported by NSF (DMR-1157490) and the State of Florida. We would like to thank the late Boris Yakshinskiy for Rutherford backscattering spectroscopy measurements, Hongwoo Baek for assisting with QHE measurement and Justin Griffith for assisting with ARPES measurement.

# Supporting Information

**Content:**

I. Methods:

Films were grown on 10 mm × 10 mm Al$_2$O$_3$(0001) substrates using custom built SVTA MOS-V-2 MBE system with base pressure of $2 \times 10^{-10}$ Torr. Substrates were cleaned ex situ by 5 minutes exposure to UV-generated ozone and in situ by heating to 750 ºC in oxygen pressure of $1 \times 10^{-6}$ Torr for ten minutes. 99.999% pure elemental bismuth, indium and selenium sources were thermally evaporated using Knudsen cells for film growth. Source fluxes were calibrated in situ by quartz crystal micro-balance (QCM) and ex situ by Rutherford backscattering spectroscopy. The ratio of selenium flux to combined bismuth and indium flux was maintained at above 10:1 as determined by QCM. For 20 QL (Bi$_{0.5}$In$_{0.5}$)$_2$Se$_3$ growth, bismuth and indium were co-evaporated by opening both shutters simultaneously, while the selenium shutter was kept open at all times during growth. For capped films, Se and MoO$_3$ were thermally evaporated at room temperature for capping.

Films were transferred from the growth chamber to an ex situ cryostat for transport measurements keeping exposure to less than 5 minutes. Magneto-resistance ($R_{XX}$) and the Hall resistance ($R_{XY}$) measurements were carried out using pressed indium leads in van der Pauw geometry in a liquid He cryogenic system with base temperature of 1.5 K and in perpendicular magnetic field ($B$) up to ± 9 Tesla. The data was symmetrized with respect to $B$ to eliminate unwanted mixing of the $R_{XX}$ and $R_{XY}$ components. The sheet carrier density was calculated from $R_{XY}$ using the Hall formula $n_{sheet} = (e\, dR_{XY}/dB)^{-1}$, where $dR_{XY}/dB$ was taken from the linear part of $R_{XY}$ for $|B| \lesssim 0.5$ T and $e$ is the electronic charge. The carrier mobility ($\mu$) was then calculated using $\mu = (e\, R_{sheet}\, n_{sheet})^{-1}$, where $R_{sheet} = R_{XX}(0)\, \pi/\ln(2)$ is the zero field sheet resistance.

For ARPES and STM measurements, the films were capped by a 100 nm selenium over-layer. For STM measurement, ion-milling was performed to remove a few nanometres of ambient contaminated selenium layer followed by annealing to ~200 ºC to evaporate rest of selenium in the STM chamber.[1] STM measurements were carried out at 78 K using Omicron LT-STM with base pressure of $1 \times 10^{-11}$ Torr. For ARPES measurement, the selenium over-layer was removed by heating the films to ~250 ºC in the ARPES chamber. ARPES measurements were then performed at room temperature using a 7 eV photon energy LASER source and a SPECS Phoibos 225 hemispherical electron analyzer.

TEM samples were prepared by focused ion beam with final $Ga^+$ ion energy of 5 keV. A JEOL ARM 200CF equipped with a cold field-emission gun and double spherical-aberration correctors operated at 200 kV were used for HAADF-STEM image acquisition with a range of detection angles from 68 to 280 mrad.

TDMTS measurements were performed in transmission geometry using a home-built THz detector. The experimental setup[2] is the same as outlined in reference 2.

For QHE measurement, an 8 QL thick $Bi_2Se_3$ film on BIS-BL was grown on 5 mm × 5mm $Al_2O_3$(0001) substrate and in situ capped by both 50 nm $MoO_3$ and 50 nm Se to prevent environmental contamination during transportation to the National High Magnetic Field Lab in Florida. The film was then hand-patterned into a Hall-bar just before measurement using a metal mask and a pair of tweezers. Hall and longitudinal resistances were measured using a Keithley 2400 source meter combined with a Keithley 7001 switch matrix in six-terminal geometry.

**II. Evaporation of $Bi_2Se_3$ through $In_2Se_3$ after 600 °C annealing during BIS-BL growth**

In order to study the effect of annealing $Bi_2Se_3$ – $In_2Se_3$ heterostructure during BIS-BL growth, we grew a 50 QL $Bi_2Se_3$ – 10 QL $In_2Se_3$ heterostructure film on $Al_2O_3$ and annealed it to 600 ºC. The sample was then analyzed using Rutherford backscattering spectroscopy (RBS), which is a quantitative tool to study thickness and composition of thin films. Figure S1 shows the data from RBS measurement along with a simulated fit obtained from SIMNRA program from which we extracted the total number of Bi, In and Se atoms in the film. From the total number of each species and known values of the atomic areal number density, we see that the film was composed of ~10 QL $(Bi_{0.02}In_{0.98})_2Se_3$ with 1% error bar in composition. This indicates that the 50 QL $Bi_2Se_3$ layer evaporated away almost entirely leaving behind an intact $In_2Se_3$ layer as shown in inset of Fig. S1. This independently confirms the results from HAADF-STEM results in Fig. 1b that shows that during the BIS-BL growth, the conducting 3 QL $Bi_2Se_3$ seed layer evaporates almost completely through the 20 QL $In_2Se_3$ layer after 600 ºC annealing, thus making the BIS-BL fully insulating.

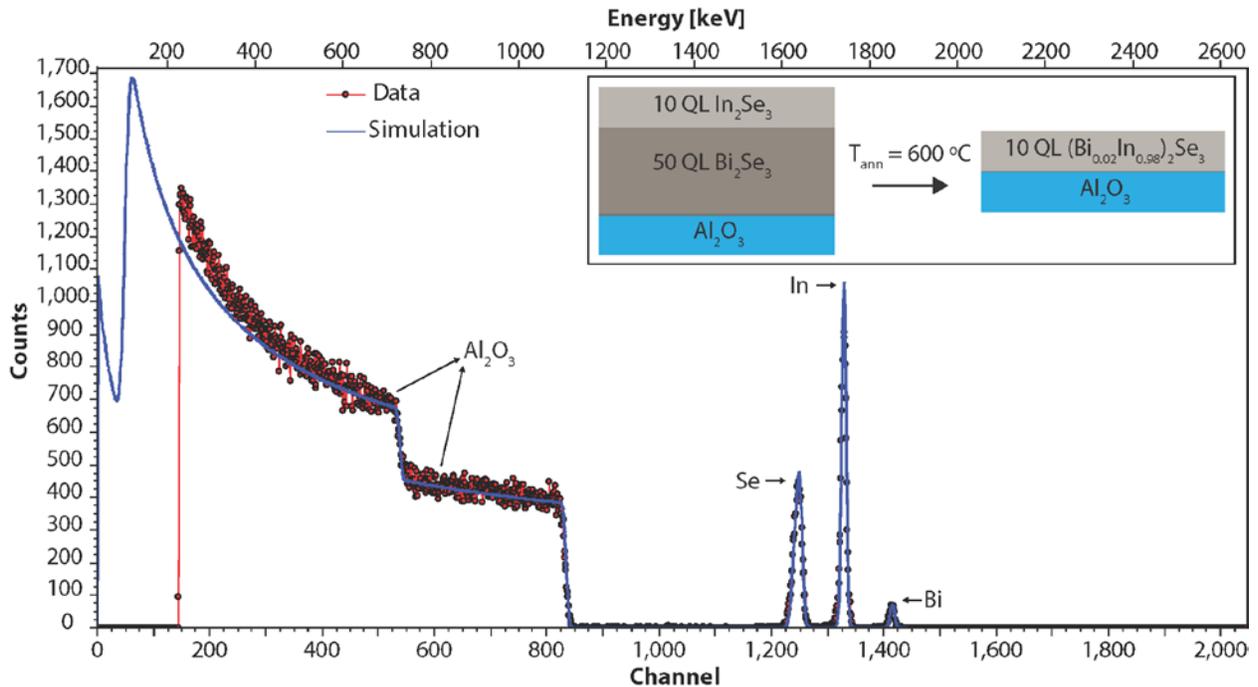

**Figure S1.** Rutherford backscattering spectroscopy (RBS) measurement on a 50 QL $Bi_2Se_3$ – 10 QL $In_2Se_3$ heterostructure annealed to 600 °C. RBS fit to the data gives ~10 QL thick $(Bi_{0.02}In_{0.98})_2Se_3$ indicating the $Bi_2Se_3$ evaporates away almost completely after annealing the heterostructure to 600 °C. The inset shows the cartoon of the film before and after the 600 °C annealing process.

### III. Nonlinear Hall effect and two-carrier model fitting

As mentioned in the main text, the non-linearity in the Hall Effect was observed at fields higher than ~0.5 T for all films, which usually indicates multiple conduction channels with different mobilities. Figure S2a shows Hall effect for 5, 25 and 60 QL thick films up to a magnetic field of 9 T. Except for the 5 QL thick film, which shows a weak non-linearity, all the other films show pronounced non-linearity similar to 25 and 60 QL thick films. For the non-linear Hall effect, the sheet carrier density calculated from low field Hall slope gives a mobility-weighted-average of different carrier species rather than carrier density of any single species. In order to specify the sheet carrier density and mobility of individual species we have used the two-carrier model (equation (S1)) to fit the Hall Effect data,

$$R_{Hall}(B) = -\frac{B}{e}\left[\frac{(n_1\mu_1^2 + n_2\mu_2^2) + B^2\mu_1^2\mu_2^2(n_1+n_2)}{(n_1\mu_1+n_2\mu_2)^2 + B^2\mu_1^2\mu_2^2(n_1+n_2)^2}\right] \quad (S1)$$

where $R_{Hall}(B)$ is the Hall resistance, $B$ is the applied magnetic field, $e$ is the electronic charge and $n_i$ and $\mu_i$ are the sheet carrier density and mobility, respectively, of $i^{th}$ species with $i = 1, 2$. $n_i$ and $\mu_i$ are the fitting parameters. Experimentally, there are only two independent parameters: we have fixed $R_{Hall}(0)/B$ to the low field slope of the Hall effect, where locally $R_{Hall}$ was linear. We have also used the zero field sheet resistance ($R_{sheet}$) = $1/[e(n_1\mu_1 + n_2\mu_2)]$ to provide an additional constraint to the fitting. This reduces the number of independent fitting parameters to just two. The model fits very well to the data as is shown in inset of Fig. S2a for a 25 QL thick film.

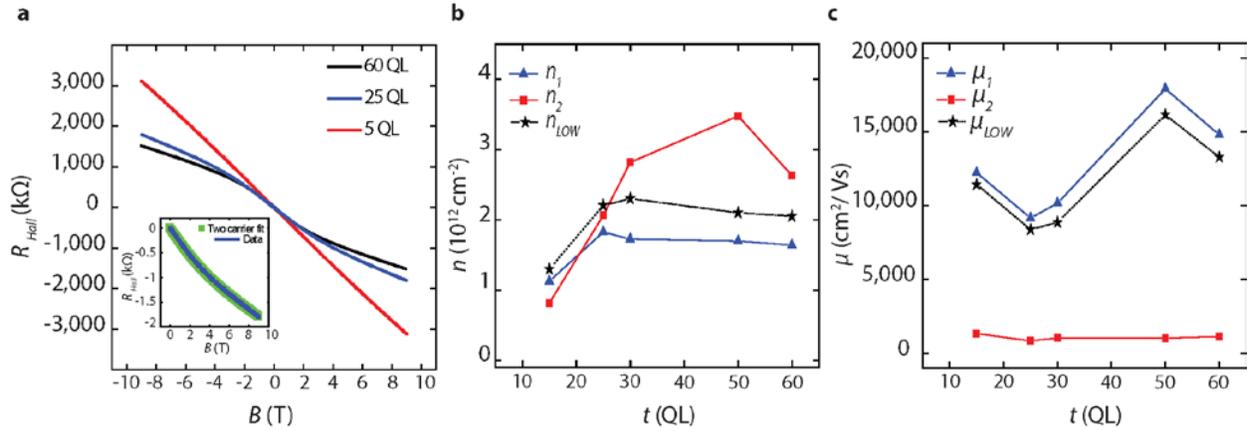

**Figure S2.** Two carrier model fitting to nonlinear Hall effect. (a) Non-linear Hall effect in 5, 25 and 50 QL thick $Bi_2Se_3$; inset shows two carrier fit to the Hall effect for 25 QL thick sample. (b) Sheet carrier density and (c) mobility obtained from two-carrier fit as a function of film thickness, for comparison low field Hall effect data are also shown.

Figure S2b shows the sheet carrier densities ($n_1$ and $n_2$) obtained from the fit along with the sheet carrier density ($n_{LOW}$) obtained from the low field Hall effect for comparison. $n_1$ clearly shows negligible thickness dependence as is the case for $n_{LOW}$. $n_2$ shows sample-to-sample variation but is essentially within ~2 - 4 × $10^{12}$ cm$^{-2}$ for films thicker than 15 QL with no clear thickness dependence within the thickness range of measured samples. In Fig. S2c, we show corresponding mobilities $\mu_1$ and $\mu_2$ for the two channels along with the low field mobility ($\mu_{LOW}$) for comparison. $\mu_1$ is comparable but consistently higher than $\mu_{LOW}$ as is expected since $\mu_{LOW}$ gives the weighted mobility of both channels rather than that of the high mobility channel alone. The fact that $n_{LOW}$ ($\mu_{LOW}$) is very close to $n_1$ ($\mu_1$) suggests that the conduction is dominated by this high mobility channel.

## IV. ARPES and estimate of In diffusion from STM

One important issue during the growth of $Bi_2Se_3$ on BIS-BL is the possibility of indium diffusion into $Bi_2Se_3$ film. It is known that there can be inter-diffusion between indium and bismuth in $Bi_2Se_3$ and $In_2Se_3$ heterostructures,[3] and the solid solution of $(Bi_{1-x}In_x)_2Se_3$ goes through a topological phase transition at $x \approx 0.03$-$0.07$, becoming a band insulator for $x > 0.25$.[4,5] The choice of $(Bi_{0.5}In_{0.5})_2Se_3$, rather than $In_2Se_3$, as the topmost layer of BIS-BL helps to minimize indium diffusion into $Bi_2Se_3$ and maintain its non-trivial topology. The TI nature of $Bi_2Se_3$ films grown on BIS-BL even in ultrathin regime is shown by observation of gapped TSS for a 5 QL thick $Bi_2Se_3$ as is indicated by the ARPES image in Fig. S3a. Such gapped TSS have been observed in ultra-thin $Bi_2Se_3$ grown on 6H-SiC (0001) with the gap attributed to hybridization of top and bottom TSS.[6] This is direct evidence of non-trivial nature of $Bi_2Se_3$ grown on top of BIS-BL even in the ultrathin limit. Such an observation means that the In diffusion should be much less than ~3%, where a signature of the topological phase transition starts to appear.[4] In the following, we give an estimate of such diffusion.

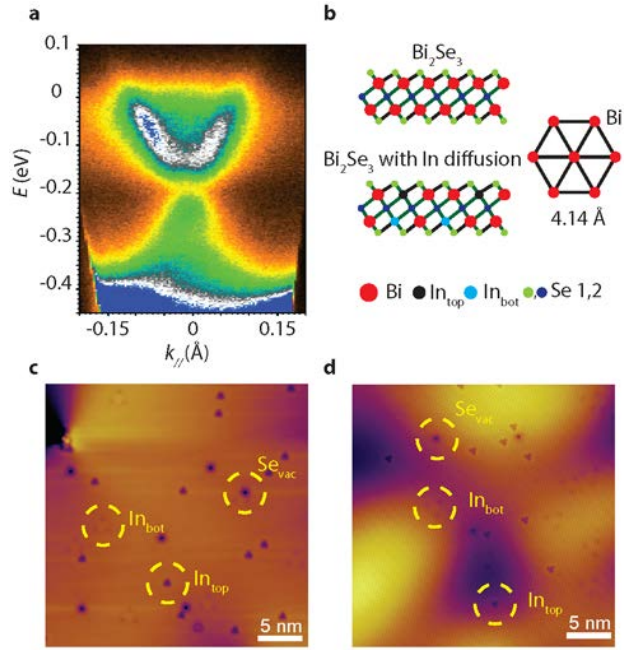

**Figure S3.** ARPES and STM of $Bi_2Se_3$ grown on BIS-BL. (a) ARPES indicating gapped TSS on 5 QL thick $Bi_2Se_3$. (b) cartoon of a QL of $Bi_2Se_3$ along with hexagonal arrangement of Bi atoms within a single layer. (c-d) STM topography image showing $In_{top}$, $In_{bot}$ and $Se_{vac}$ defects on the topmost QL of (c) 30 QL and (d) 5 QL thick $Bi_2Se_3$

Figure S3b shows a cartoon of a quintuple layer (QL) of $Bi_2Se_3$, which consists of alternate layers of Se-Bi-Se-Bi-Se. For $Bi_2Se_3$ grown on BIS-BL some In diffusion from BIS-BL to $Bi_2Se_3$ is observed, which is shown schematically in Fig. S3b (bottom left image). Diffused In atoms preferentially occupy the Bi sites in the upper and lower Bi layers within a QL and are denoted as $In_{top}$ and $In_{bot}$, respectively. Figures S3c and S3d show scanning tunneling microscopy (STM) images of 30 QL and 5 QL $Bi_2Se_3$ films grown on BIS-BL respectively with three kinds of defects identified – Se vacancy ($Se_{vac}$), $In_{top}$ and $In_{bot}$.

By counting $In_{top}$ and $In_{bot}$ on the surface and calculating the areal number density of Bi ($\rho_{Bi}$), we can estimate the percentage of In diffusion ($In\%$) in $Bi_2Se_3$. In order to calculate $\rho_{Bi}$, we note that Bi atoms are arranged in hexagonal pattern within a monolayer of Bi. A schematic of such a hexagonal unit for Bi layer is shown on Fig. S3b. Since the in-plane lattice constant of Bi$_2$Se$_3$ is ~4.14 Å, the area of the hexagonal unit is ~4.45 × 10$^{-15}$ cm$^2$. There are 3 Bi atoms in one hexagonal unit, $\rho_{Bi} \approx 6.74 \times 10^{14}$ cm$^{-2}$. From the STM image in Fig. 3d we can count ~11 $In_{top}$ and ~9 $In_{bot}$ in a 30 nm × 30 nm area (*area*). This gives $In\%$ = ($In_{top}+In_{bot}$)/(*area*×$2\rho_{Bi}$)×100% on the topmost QL of Bi$_2$Se$_3$ to be ~0.2% for 5 QL Bi$_2$Se$_3$, where 2 in the denominator accounts for two Bi layers within a QL. Similar calculations for a 30 QL thick film also give ~0.2% In diffusion.

## V. Consistency of transport data and ARPES with TSS conduction

From Hall measurement, it is clear that low field sheet carrier density ($n_{low}$) is less than ~2 × 10$^{12}$ cm$^{-2}$ in the entire thickness range. Two-carrier fitting from the Hall effect measurement gives a total sheet carrier density ($n_{tot} = n_1 + n_2$) to be at most ~5 × 10$^{12}$ cm$^{-2}$. In Bi$_2$Se$_3$, when the total sheet carrier density is ~1 × 10$^{13}$ cm$^{-2}$ (or equivalently ~5× 10$^{12}$ cm$^{-2}$ per surface) the surface Fermi energy lies at the bottom of the bulk conduction band[7]. Given that the total carrier density $n_1 + n_2$, is much smaller than ~1 × 10$^{13}$ cm$^{-2}$ for films grown on BIS-BL, they should have, if anything, upward band bending resulting in formation of a depletion region. Such upward band bending cannot form quantum well states or 2DEG[7,8]. Therefore, the most consistent interpretation of the observed channels with thickness independent sheet carrier density is that both of them originate from the TSS. This is also supported by ARPES data (Fig. 2c in the main text), where the surface $E_F$ lies in the bulk band gap and no such 2DEGs are observed. In contrast ARPES measurements on Bi$_2$Se$_3$ grown directly on Al$_2$O$_3$ clearly show presence of such 2DEG states (Fig. 2d in the main text). In order to show the existence of a 2DEG state is unlikely to be present, we can estimate the expected sheet carrier density of TSS if either $n_1$ or $n_2$ originates from 2DEGs. Let us assume that $n_1 \approx 1.8 \times 10^{12}$ cm$^{-2}$ is due to 2DEG carriers. Then we can get the Fermi wave-vector for 2DEG using $k_{F,2DEG} = \sqrt{2\pi n_1}$. This results in $k_{F,2DEG} = 0.034$ Å. Using the ARPES spectrum of $Bi_2Se_3$ grown on $Al_2O_3$, we can extrapolate the Fermi wave-vector of corresponding TSS ($k_{F,TSS}$) at this $k_{F,2DEG}$.

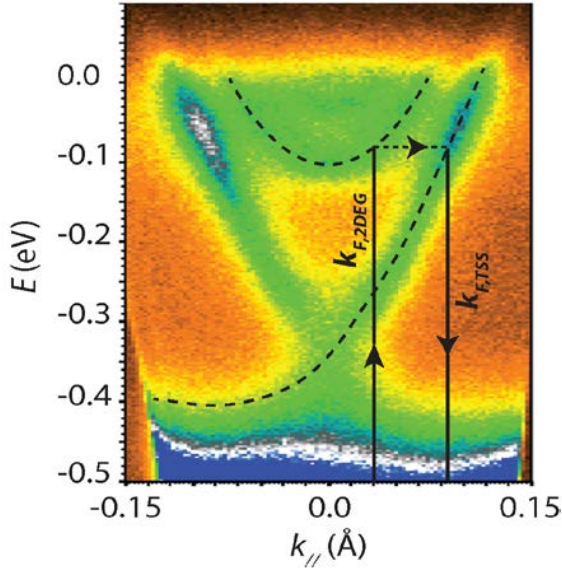

**Figure S4.** Extrapolation of TSS wave vector from that of 2DEG using ARPES of $Bi_2Se_3$ grown on $Al_2O_3$.

As shown in Fig. S4, such extrapolation yields $k_{F,TSS} \approx 0.088$ Å. From $k_{F,TSS}$ we can get $n_{sheet,TSS} = k_{F,TSS}^2/4\pi \approx 6.1 \times 10^{12}$ cm$^{-2}$ for corresponding TSS, where the 4 in the denominator is due to the non-degenerate nature of TSS. Similar estimation assuming $n_2 \approx 3 \times 10^{12}$ cm$^{-2}$ to come from 2DEG yields $n_{sheet,TSS} \approx 6.8 \times 10^{12}$ cm$^{-2}$. This gives a combined TSS and 2DEG sheet carrier density of ~$7.9 \times 10^{12}$ cm$^{-2}$ (~$9.8 \times 10^{12}$ cm$^{-2}$) from a single surface assuming $n_1$ ($n_2$) originates from 2DEG state. For simplicity if we assume the other surface to have similar carrier density, then the total carrier density would be well above ~$10^{13}$ cm$^{-2}$, which is over three times that of what is observed from Hall effect which rules out the presence of 2DEGs. Therefore, it is most natural to associate the two channels to the TSSs from the top and bottom surfaces, respectively. Naturally, the following question arises: which of the two TSSs is responsible for the higher mobility channel? This can be indirectly answered from the capping layer samples. Considering that the mobilities of the Se and $MoO_3$ capped films are substantially reduced from uncapped samples, it seems that the higher mobility channel originates from the top TSS; if the high mobility channel originated from the bottom TSS, such dramatic reduction would not be expected with capping.

Finally, we compare the Fermi level ($E_F$) obtained from Hall effect to that obtained from ARPES on 30 QL $Bi_2Se_3$ grown on BIS-BL, which is shown on Fig. 3a of the main text. From ARPES, $E_F$ is observed to be ~0.17 eV above the Dirac point and the Fermi wave-vector ($k_F$) is observed to be ~0.052 Å$^{-1}$. In order to obtain $E_F$ from Hall effect, we can calculate $k_{F,Hall}$ from $k_{F,Hall} = \sqrt{4\pi n_1}$, where $n_1$ is obtained from two-carrier fit of Hall Effect measurement and the pre-factor 4 is due to spin non-degenerate nature of TSS. Since $n_1$ is ~$1.8 \times 10^{12}$ cm$^{-2}$ for the entire thickness range, this gives $k_{F,Hall}$ to be ~0.0475 Å$^{-1}$. Using the ARPES image shown in Fig. S4, this $k_F$ corresponds to $E_F \approx 0.14$ eV, which is slightly smaller than $E_F \approx 0.17$ eV that is directly observed from ARPES. Such a difference could be due to the sample to sample variation or due to the high temperature annealing (~250 ºC) required in Se decapping process during ARPES measurement, which could lead to creation of more Se vacancies, therefore pushing the Fermi level higher.

**VI. SdH oscillation vs. Cyclotron Resonance**

Cyclotron mass can be obtained from Shubnikov-de Haas oscillations in magnetoresistance measurement or from cyclotron resonance in magneto-optical measurement. Despite the significantly enhanced Hall mobilities, no Shubnikov–de Haas (SdH) oscillations were observed in fields up to 9 Tesla. This is surprising considering that the standard $Bi_2Se_3$[9] or even Cu-doped

Bi$_2$Se$_3$[7] films, having much lower mobilities exhibit well developed SdH oscillations in fields higher than ~5 T. Although the origin for the absence of SdH oscillations in these high mobility films is not fully understood yet, one possibility is due to the carrier density inhomogenity that may be more severe in these low carrier density samples.[10] As previously pointed out in conventional two dimensional electron gas system of similar sheet carrier densities in GaN/AlGaN hetereostructures[11] any slight inhomegeneities in the carrier density can significantly suppress the SdH oscillation due to the phase cancelling effect: this view is further supported by the very observation of the full quantum Hall effect when all carriers are driven to the lowest Landau level, where the effect of any inhomogeneity in carrier densities vanishes. In contrast, well developed cyclotron resonance was observed in time domain magneto-terahertz spectroscopy measurement, from which cyclotron mass was extracted.

**VII. Time domain magneto-terahertz spectroscopy measurement and fits**

In order to measure the complex Faraday rotation (FR), we used the phase modulation technique to measure the polarization states accurately which allows us to measure $E_{xx}(t)$ and $E_{xy}(t)$ simultaneously in a single scan.[2] Faraday rotation can be obtained by $\theta_F = arctan(E_{xy}(\omega)/E_{xx}(\omega))$ = $\theta_F' + i\,\theta_F''$ after Fourier transforming into the frequency domain. Sapphire has no detectable FR and 20 nm Se or 50 nm MoO$_3$ do not show rotation within our experimental accuracy (0.5 mrad). We subtracted the non-smooth background from a reference substrate before fitting the data. Field dependent imaginary parts of the complex FR are shown in Fig. S5a and S5b for Se-capped and MoO$_3$-capped film respectively. The imaginary part of FR is related to the ellipticity.[2] Corresponding real parts are shown in main text (Fig. 3a and 3b).

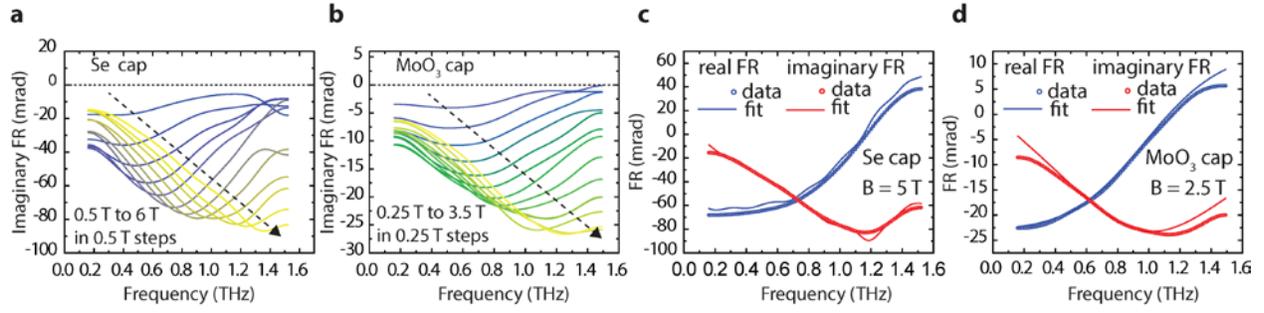

**Figure S5.** Faraday rotation at different magnetic fields and representative fit for Se and MoO$_3$ capped film. (a-b) Dip-like features in imaginary part of the FR that shift to higher frequency with increasing magnetic field indicate the cyclotron resonance in (a) 20 nm Se-capped film and (b) 50 nm MoO$_3$-capped film. The dashed arrow shows the direction of increasing magnetic field in steps as defined in figures. (c-d) Representative fit for imaginary and real part of FR at magnetic field of 5 T for (c) Se-capped film and 2.5 T for (d) MoO$_3$-capped film.

The dip position in the imaginary part indicates cyclotron resonance (CR) frequency. The negative Faraday rotation indicates that carriers are electrons in both films. We fit the data by Drude-Lorentz model with a Drude term, a phonon term and a term for the background dielectric constant ($\epsilon_\infty$) coming from higher energy absorptions. The formula for conductance in magnetic field is

$$G_\pm = -i\epsilon_0 \omega d \left[ \frac{\omega_{pD}^2}{-\omega^2 - i\Gamma_D \omega \mp \omega_c \omega} + \frac{\omega_{pDL}^2}{\omega_{DL}^2 - \omega^2 - i\Gamma_{DL}\omega \mp \omega_{cDL}\omega} + \epsilon_\infty - 1 \right]$$

where $\omega_p$ represents the plasma frequencies, $\Gamma$ represents scattering rates, $d$ is the film thickness and the ± sign denotes the response to right/left circularly polarized light respectively. We constrained the parameters of the phonon and the high-frequency terms by those extracted from zero-field conductance value (as explained below) and only allowed the cyclotron frequency ($\omega_c$) and the scattering rate to vary. From $G_\pm$, we can calculate the complex transmission for right and left circularly polarized light $t_\pm$. Then we can calculate the complex FR by $tan(\theta_F) = -i(t_+ - t_-)/(t_+ + t_-)$. From the fits we can accurately extract the cyclotron frequency, $\omega_c$, for the Drude component from which the cyclotron mass ($m^*$) is calculated using $\omega_c = eB/(2\pi m^*)$. Figures S5c and S5d show fits to the real and imaginary part of Faraday rotation for Se-capped film at 5 T and MoO$_3$-capped film at 2.5 T, respectively. Similarly, we fitted the zero field real conductance spectra by an oscillator model with a Drude term describing free electron-like motion, a Drude-Lorentz term modeling the phonon and a lattice polarizability ($\epsilon_\infty$) term that originates from absorptions outside the spectral range.

$$G(\omega) = \left[ -\frac{\omega_{pD}^2}{i\omega - \Gamma_D} - \frac{i\omega \omega_{pDL}^2}{\omega_{DL}^2 - \omega^2 - i\omega\Gamma_{DL}} - i(\epsilon_\infty - 1) \right] \epsilon_o d$$

$\omega_{pD}$, $\omega_{pDL}$, $\Gamma_{DL}$, and $\epsilon_\infty$ obtained from this fit were used to constrain the fit at finite magnetic field.